Submitted to **ADVANCED MATERIALS**

DOI: 10.1002/adma.((please add manuscript number))

# Strain and magnetic field induced spin-structure transitions in multiferroic BiFeO$_3$

By A. Agbelele,[†] D. Sando,[†*] C. Toulouse, C. Paillard, R. D. Johnson, R. Rüffer, A. F. Popkov, C. Carrétéro, P. Rovillain, J.-M. Le Breton, B. Dkhil, M. Cazayous, Y. Gallais, M.-A. Measson, A. Sacuto, P. Manuel, A. K. Zvezdin, A. Barthélémy, J. Juraszek, and M. Bibes[*]

[*]     Dr. D. Sando, Dr. C. Carrétéro, Prof. A. Barthélémy, Dr. M. Bibes
Unité Mixte de Physique, CNRS, Thales, Univ. Paris-Sud, Université Paris-Saclay, 91767 Palaiseau, France
Email: daniel.sando@unsw.edu.au
manuel.bibes@thalesgroup.com

Dr. A. Agbelele, Prof. J.-M. Le Breton, Dr. J. Juraszek
Groupe de Physique des Matériaux, UMR 6634, CNRS-Université de Rouen, Avenue de l'Université BP12 Saint Etienne du Rouvray Cedex, France

Dr. D. Sando
School of Materials Science and Engineering, UNSW Australia, Sydney, NSW 2052, Australia

C. Toulouse, Dr. P. Rovillain, Dr. M. Cazayous, Dr. Y. Gallais, Dr. M.-A. Measson, Prof. A. Sacuto
Laboratoire Matériaux et Phénomènes Quantiques (UMR 7162 CNRS), Université Paris Diderot-Paris 7, 75205 Paris Cedex 13, France

C. Paillard, Prof. B. Dkhil
Laboratoire SPMS, UMR 8580, Ecole Centrale Paris-CNRS, Grande Voie des Vignes, Châtenay-Malabry, France

Dr. R. D. Johnson
Clarendon Laboratory, University of Oxford, Parks Road, Oxford, OX1 3PU, United Kingdom

Dr. R. Rüffer
European Synchrotron Radiation Facility, CS 40220, F-38043 Grenoble Cedex 9, France

Dr. A. F. Popkov, Prof. A. K. Zvezdin
Moscow Institute of Physics and Technology (State University), 141700, Dolgoprudny, Russia

Dr. A. F. Popkov
National Research University of Electronic Technology (MIET), Pas. 4806, Bld. 5, Zelenograd, Moscow, Russia

Dr. P. Manuel
ISIS Facility, STFC, Rutherford Appleton Laboratory, Didcot OX11 0QX, United Kingdom

Prof. A. K. Zvezdin
Prokhorov General Physics Institute, Russian Academy of Sciences, 119991, Moscow, Russia

Prof. A. K. Zvezdin
Russian Quantum Center, 143025, Skolkovo, Moscow, Russia

In multiferroic materials,[1] the coexistence of several exchange interactions often results in competition between non-collinear spin orders which are sensitive to temperature, hydrostatic pressure, or magnetic field. In bismuth ferrite ($BiFeO_3$), a room-temperature multiferroic,[2] the intricacy of the magnetic phase diagram is only fully revealed in thin films:[3] epitaxial strain suppresses the cycloidal spin order present in the bulk,[4] transforming it into various antiferromagnetic states, modifying the spin direction and ordering patterns.[5] Here, we explore the combined effect of strain and magnetic field on the spin order in $BiFeO_3$. Through nuclear resonant scattering[6] and Raman spectroscopy, we show that both strain and magnetic field destabilize the cycloid, resulting in a critical field sharply reduced from the bulk value. Neutron diffraction data support this hypothesis, with a cycloid period larger than the bulk value and increasing with strain and/or magnetic field. Analysis of the data in light of Landau-Lifshitz calculations[7] indicates that very small strains are sufficient to induce large modifications in magnetoelastic coupling,[8] suggesting interesting opportunities for strain- and/or field-mediated devices which take advantage of finite-size effects in multiferroic films.

Frustration in magnetic systems with interplay between spin and charge often brings about non-collinear orders such as spin spirals and cycloids.[9] These configurations can arise either directly from competing exchange interactions, as is the case for the spiral order in $TbMnO_3$,[10] or through the influence of the ferroelectric polarization on the spin arrangement (via a Dzyaloshinskii-Moriya-like interaction[11]). In such materials, perturbations such as strain[5], pressure,[12] doping,[13] or external field can markedly modify the spin structure. These stimuli may even destroy the cycloidal order, releasing a weak canted magnetic moment and allowing the linear magnetoelectric effect. To take full advantage of magnetoelectric effects and achieve efficient electrical control of magnetic order and spin excitations in such materials,[14–16] understanding these transitions is an important step.

The perovskite compound $BiFeO_3$ (BFO) is an ideal test bed for exploring such effects. It orders both ferroelectrically ($T_C$ = 1100 K) [17] and antiferromagnetically ($T_N$ = 640 K) [18] above room temperature, and can be prepared in thin films over a wide range of strain. Importantly, its magnetic ion can conveniently be substituted by Mössbauer-active $^{57}$Fe,[19] enabling detailed studies of magnetism that are generally not possible in other non-ferrous cycloidal multiferroics. BFO is presently the focus of intense research, not least of all for its fascinating physics and promise in the field of spintronics,[20] but also in realms such as optics,[21–23] domain wall nanoelectronics,[24,25] and magnonics.[15,26]

Bulk BFO possesses G-type antiferromagnetic order, upon which is superimposed an incommensurate cycloidal modulation of wavelength ~62 nm. The cycloid, which lies in the plane containing the electric polarization (typically <111>) and the <1-10> direction, is suppressed – yielding to simple G-type order – by a critical magnetic field of $H_{cr}$ = 18 T.[27–29] In our previous work,[5] we employed epitaxial strain engineering to show that the spin structure in BFO is strongly strain-dependent. Here, we explore the combined effects of strain and magnetic field on the spin order of BFO. The $^{57}$Fe substitution in our films allows us to establish the magnetic properties by exploiting the uncommon, yet powerful technique of nuclear resonant scattering (NRS) of synchrotron radiation. We find that in BFO films which display cycloidal order, the critical field is sharply reduced when compared to the bulk. Neutron diffraction experiments on one such film indicate that the cycloid period is larger by around 20 nm in this specimen than in the bulk. Complementary Raman spectroscopy measurements indicate that the transition from cycloidal to homogeneous order is accompanied by a further lengthening of the cycloid period, up to the critical field at which the cycloid is destroyed. Finally, we use Landau-Lifshitz formalism to show that the marked decrease in critical field arises from a combination of enhanced anisotropy and strong magnetoelastic effects which are inherent to the thin film geometry. Our results indicate that thin films offer a unique opportunity for exploring spin structure transitions in multiferroics and suggest interesting device opportunities making use of symmetry- and strain-induced changes in such systems.

First we consider BFO's room-temperature magnetic order as a function of strain, determined by NRS. This technique, which is the time-analogue of Mössbauer spectroscopy, probes the hyperfine splitting of the nuclear levels in $^{57}$Fe. The experiment is performed using focused 14.413 keV x-ray pulses in grazing incidence geometry, **Figure 1(a)**. The pulses simultaneously excite the $^{57}$Fe nuclei into the hyperfine-split nuclear energy levels [Figure 1(b)], which then decay to their initial state. These various decay channels interfere, producing distinct *quantum beat* patterns[6] whose shape depends on the local direction of the hyperfine field represented by polar and azimuthal angles $\theta$ and $\varphi$ [Figure 1(c)].

NRS spectra measured at 300 K for strained ~70 nm thick (001)-oriented BFO films are presented in Figure 2(a) (pseudocubic notation is used throughout). The differences in the spectra suggest significant strain-induced modifications in the magnetic order; fitting with either pseudo-collinear order or spin-cycloid models (see details in the Supporting Information), we infer the spin structures illustrated in Figure 2(b). For high compressive strain (SrTiO$_3$ - STO), the spins are distributed in an easy-plane state, while for high tensile strain (NdScO$_3$ - NSO) the spins are tilted towards the out-of-plane direction, subtending an angle of ~34° with the film normal. We point out here that in Figure 2(b), we do not imply a conical spin structure: since the NRS experiment probes multiple magnetic domains in our films, we show a possible distribution of spins compatible with the experimental data. For lower values of strain (GdScO$_3$ – GSO and SmScO$_3$ - SSO), a cycloidal order is observed: for BFO//GSO ($\varepsilon = -0.1\ \%$), the data are well-fitted using a magnetic structure consistent with the bulk-like cycloid with a propagation direction of [1-10], while for BFO//SSO ($\varepsilon = +0.2\ \%$), the best fit is obtained when considering the type-2 cycloid[5] with propagation direction along [110]. These spin structures are consistent with those obtained with Mössbauer spectroscopy measurements.[5]

We now describe the effects on the spin structure of a magnetic field applied along the [001] direction; *i.e.* normal to the film surface, focusing on the spectra for the weakly-strained films which display the cycloid. For both BFO//GSO and BFO//SSO [Figures 3(a, e)] increasing the field progressively modifies the NRS spectra. For BFO//GSO, the fitting results indicate that for fields up to 4 T the cycloidal order is preserved. Figure 3(c) shows that the values of $\theta$ as a function of $\varphi$ extracted from the fits are consistent with the relation expected for the cycloid, for which $\theta(\varphi)$ is plotted as a line. On the other hand, under an applied field of 6 T, the data are well-fit with a pseudo-collinear spin order with a distribution of spin directions in the (111) plane as plotted in Figure 3(d). These data therefore imply the occurrence of a cycloid → homogeneous order transition close to 6 T. The weighting of the spin directions in the fits for the non-cycloidal state at 6 T [Figure 2(d)] indicate that the preferred spin direction is perpendicular to the applied magnetic field.

For the BFO film grown on SSO, the type-2 cycloid[5] has spins that are modulated in a plane defined by the <110> direction and [001] [Figure 3(e)]. This spin order results in a range of $\theta$ but a fixed value of $\varphi = 45°$ (see Supporting Information). At low applied field, the NRS spectrum for BFO/SSO is consistent with the type-2 cycloid [Figure 3(g)], while at applied fields above 2 T, two orthogonal preferred directions for the Fe spins in the (111)-type plane are deduced from the fit [Figure 3(f)]: one is perpendicular to the applied field (*i.e.* in the sample plane), the other points out of the sample plane due to strain-induced anisotropy. Here we infer a cycloid $\rightarrow$ homogeneous order transition at applied field between 2 and 4 T.

Next we describe the results of low-energy Raman spectroscopy of strained BFO films under magnetic field. The Raman signature of the incommensurate magnetic cycloid in BFO is two series of sharp peaks measured at low energy and selected with different polarization of the incident and scattered light. The peaks correspond to magnon modes associated to the spin oscillations in and out of the cycloidal plane – labelled $\Phi$ and $\psi$ modes respectively. This signature originates from the translational symmetry breaking of the cycloidal ground state.[30,31] Figure 4 presents the evolution of the low energy Raman signal for strained BFO films grown on $DyScO_3$ (DSO), GSO, and SSO under a magnetic field applied along [001]. Upon increasing field, the $\Phi$ modes transition from multiple peaks to a single peak, signalling the transition from non-collinear to a homogeneous magnetic state, *i.e.* the destruction of the cycloid. The critical field for this transition is strain-dependent: the modification of the spin order occurs at ~4 T for BFO/DSO, above 6 T for BFO/GSO, and above 2 T for BFO/SSO. For fields above these values, the magnetic order can be described as a simple two-sublattice antiferromagnet.

The NRS and Raman experiments both indicate that in the thin film geometry the spin cycloid is dramatically destabilized and give comparable values of critical field to induce collinear order. Moreover, films under a larger strain exhibit the lowest critical field for the transition to the homogenous state, *cf.* Figure 5(a). It is interesting to note that a lower critical field suggests a lower cycloid energy and thus a cycloid with longer period, which has been observed in doped bulk BFO,[13,32] in thin films under strain[33] and in single-crystal BFO when $T_N$ is approached.[34] To explore the possibility of a longer cycloid period in our films, we turned to neutron diffraction.

Single crystal neutron diffraction on BFO//GSO (misfit strain of -0.1 %) was performed using WISH,[35] at ISIS, UK. Diffraction data were collected at room temperature, and a single, broad magnetic peak was observed at the structurally forbidden (½ ½ ½) position in reciprocal space. Diffraction from a single cycloidal magnetic domain gives rise to a pair of satellite peaks of equal intensity about the (½ ½ ½) reflection, with the separation of the satellites being inversely proportional to the real-space periodicity of the magnetic cycloid. In this experiment one would expect to observe three pairs of satellites which correspond to three cycloidal domains related by the three-fold symmetry of the pseudo-cubic [1 1 1] axis; however, no satellites were resolved. We can, however, ascertain a lower limit on the cycloid periodicity, as described in the following. Figure 5(b) presents the magnetic diffraction intensity integrated along the (-1 1 0) direction. (Note that the point density in the figure gives an accurate representation of the instrumental resolution.) Narrow limits in d-spacing centred on the (½ ½ ½) satellites were employed to filter out diffraction from other minority ferroelastic domains. Diffraction peaks from all three domains were fit to the 1-D data assuming equal domain populations. The length of the cycloid was free to refine, giving a period of 82.2 ± 7.3 nm – some 20 nm (about 30%) larger than that observed in bulk BFO. This result was cross-checked against analysis of the same diffraction intensity, but instead focused in *d*-spacing rather than reciprocal lattice units. Again, an extended period cycloid was found for the film.

These diffraction data thus offer some hints as to why we observe a much lower critical field in our thin films than in the bulk: a larger cycloid period may imply the existence of an extra uniaxial anisotropy term, thereby lowering the energy (and thus field) required to destroy the cycloidal order. Taking the cycloid period for BFO//GSO at 0 T to be 82.2 ± 7.3 nm, and using the peak period from the Raman data at 0 T [Figure 5(c)], we deduce a magnon velocity $(1.57 ± 0.20) \times 10^4$ m.s$^{-1}$; rather close to the value measured in bulk BFO.[36] Then, assuming a constant magnon velocity in all samples at all fields, we deduce the cycloid period under field for our three samples as shown in Figure 5(d). It can be seen that the cycloid period exhibits an increase upon approach to the critical field; this is particularly evident for BFO//GSO for which the critical field is the highest. The guide lines use the form of cycloid period lengthening under magnetic field proposed by Gareeva et al.[7]

To delve deeper into the origin of the decreased critical field, we use a Laundau-Lifshitz formalism for the free energy of the system (the details of which are described in the Experimental Section). In this treatment, we take into account the inhomogeneous exchange energy, macroscopic Dzyaloshinskii-Moriya interaction (responsible for the cycloidal ordering), Zeeman energy for the effective field (the external applied field and internal fields related to the weak ferromagnetism), and the uniaxial anisotropy which includes a magnetoelastic term taking into account the epitaxial strain.

Most important in the thin-film case is the magnetoelastic term: In order to understand the dramatically reduced critical field we must consider that the magnetoelastic coupling coefficients for BFO thin films may be vastly different (sometimes even in sign) from their bulk values. To reproduce the reduced critical field in our films, we consider an effective magneto-elastic coupling coefficient that follows a quadratic strain dependence.[8]

The results of our theoretical analysis of the cycloidal modulation in (001)-oriented films are summarized in Figure 5(a), where it is shown that the critical field for cycloid suppression is strongly dependent on strain. We point out that the theoretical curve is valid for the type-1 (bulk-like) cycloid only; the type 2 cycloid observed for slight tensile strain[5] has a critical field lower than for the type-1 cycloid, explaining the slight discrepancy in the agreement shown in Figure 5(a).

The implications of the results shown in Figure 5(a) are manifold. First, they imply that even in the absence of mismatch strain, other factors such as changes in symmetry[5,37] and finite size effects which occur in thin films, reduce the critical field by a factor of two relative to bulk BFO. Second, our calculations allow us to estimate the magnetoelastic energy in strained BFO films (see Figure S5), an important parameter for engineering of magnetic order in thin films. Taken as a whole, these observations imply that the linear magnetoelectric effect in BFO is accessible by a combination of epitaxial strain and moderate applied fields, suggesting interesting device opportunities using strained BFO films.

In summary, we have explored the effect of magnetic field on the spin order in strained $BiFeO_3$ films. Using nuclear resonant scattering, Raman spectroscopy, and Landau-Lifshitz theory, we have shown that the critical field to suppress the cycloidal ordering is strongly reduced when compared with bulk BFO. Calculations indicate that the magnetoelastic coupling parameters in our films are dramatically different from their bulk values, resulting in markedly different field-dependent spin dynamics. We also find that as the field approaches the critical value, the cycloid period is sharply accentuated. As well as facilitating access to the linear magnetoelectric effect at moderate values of applied field, our results show that strain engineering and finite-size effects exploited in parallel provide a powerful approach for inducing novel magnetic phases in thin-film multiferroics. In particular, the strong modification of the spin wave mode frequencies for moderate applied fields is a feature that may be attractive for magnonic devices.[26]

**Experimental Section**

***Thin film growth*** of strained epitaxial films of BiFeO$_3$ was performed by pulsed laser deposition on various single crystal substrates using conditions described previously [38]. Typical film thickness was 70 nm. Details of the structural properties of such films can be found in Refs. [5,39]. Targets were ~100 % enriched in $^{57}$Fe to enhance the signal in NRS measurements.

***Nuclear Resonant Scattering*** (NRS) in grazing incidence geometry was carried out at the Nuclear Resonance Beamline (ID18) [40] at ESRF, Grenoble, France. Magnetic fields up to 6 T were applied and all measurements described here were taken at room temperature. The recorded NRS time spectra were fitted using the CONUSS program[41] (more details can be found in the Supporting Information). For non-collinear spin states, we used ten orientations of the magnetic hyperfine field following a theoretical cycloidal spin distribution (with $\theta$ and $\varphi$ as free parameters). The hyperfine field was slightly distributed around 48.4 T to take into account an anisotropy of the hyperfine interaction, and the quadrupole splitting was fixed to the bulk BFO value of 0.44 mm·s$^{-1}$. For collinear spin states, we used a model based on (111) magnetic easy-plane [42] for the hyperfine field orientations to fit the NRS spectra. In this case, the relative weight of each possible easy-axis was fitted for each spectrum allowing us to deduce the Fe spin distribution in this plane.

***Raman spectroscopy*** was performed in the backscattering geometry using a 647.1 nm laser line. Raman scattering was collected by a triple spectrometer (Jobin Yvon T64000) equipped with a charge-coupled device (CCD). The spot size was about 100 μm$^2$ and the penetration depth was less than 100 nm. The $\varphi_n$ modes, where the *n* index labels the modes from their lowest to highest energy, were selected using parallel polarizers in the (010) plane.

Measurements under magnetic field up to 8 T were obtained using an Oxford Spectromag split-coil magnet.

*Neutron diffraction* was performed at WISH, a time-of-flight diffractometer at ISIS, UK. The sample was mounted with the (111) reflection in the horizontal scattering plane and with $2\theta \approx 140°$, chosen to optimize instrument resolution and incident flux at the respective wavelength.

*Landau-Lifshitz Calculations*: To analyse homogeneous and spatially-modulated antiferromagnetic states in $BiFeO_3$ grown on a substrate in the (001) orientation, we used the approach of quantitative and qualitative analysis of reduced Landau-Lifshitz (LL) equations for the antiferromagnetic moment, similar to that described in [7] for the bulk multiferroic. In the present case, the reduced LL equations are obtained by the variation of the thermodynamic potential of the multiferroic. This includes the energy of inhomogeneous exchange and magnetoelectric interactions, the energy of interaction with the external magnetic field, and the magnetic anisotropy energy, which in turn consists of the bulk anisotropy due to the Dzyaloshinskii-Moria interaction and uniaxial anisotropy induced during film growth:

$$F(\mathbf{l}) = A\left[(\nabla l_x)^2 + (\nabla l_y)^2 + (\nabla l_z)^2\right] + \beta \mathbf{e}_p\{(\mathbf{l} \cdot \nabla)\mathbf{l} - l(\nabla \cdot \mathbf{l})\} + \frac{\chi_\perp}{2}H_D^2(\mathbf{e}_p \cdot \mathbf{l})^2 + \frac{\chi_\perp}{2}(\mathbf{Hl})^2 -$$

$$-\chi_\perp H_D \mathbf{H} \cdot [\mathbf{e}_c \times \mathbf{l}] + K_{111}(\mathbf{e}_p \cdot \mathbf{l})^2 + K_{001}(\mathbf{e}_s \cdot \mathbf{l})^2.$$

Here, $\mathbf{l}$ is the unit antiferromagnetic vector, $A$ is the stiffness constant, β is the constant of the non-uniform magnetoelectric interaction, $\mathbf{e}_p$ is the unit vector of spontaneous polarization $\mathbf{P}$ oriented along the principal crystal axis [111], $\mathbf{e}_c$ is the unit vector along [111], $\mathbf{e}_s$ is the unit vector along [001], $H_D$ is the Dzyaloshinskii field, $\chi_\perp$ is the transverse magnetic susceptibility of the antiferromagnet, $K_{111}$ is the uniaxial bulk anisotropy, and $K_{001}$ is the induced anisotropy. In the calculations we used the following parameters: $= 3 \cdot 10^{-7}$ erg·cm$^{-1}$, $\chi_\perp = \frac{M_0^2}{a} = 4 \cdot 10^{-5}$; $\beta = 0.6$ erg·cm$^{-2}$, $H_D = 10^5$ Oe, $M_0 = 640$ emu·cm$^{-3}$, $K_{111} = 1.7 \cdot 10^4$ J·m$^{-3}$.

From analysis of the stability of the homogeneous state we calculate the critical magnetic field of the transition to an incommensurate antiferromagnetic phase in the energy dependence of the induced anisotropy in the film $H_C(K_{001})$. It is assumed here that the induced anisotropy for the large elastic strain arising due to the mismatch of the lattice parameters of the film and substrate materials is caused not only by a linear, but also a quadratic contribution to the magnetoelastic energy, *viz.*

$$K_{001}(\varepsilon) = (B + D\varepsilon)\left(\frac{1+\upsilon}{1-\upsilon}\right)\varepsilon,$$

where B and D are effective parameters of the magnetoelastic energy, and $\upsilon$ is Poisson's ratio [43]. This leads to a real dependence of the critical magnetic field on the magnitude and sign of the elastic deformation. So, one can obtain for example a quadratic approximation thus

$$H_m(\varepsilon) \approx H(0) + \frac{dH}{dK_{001}}K_{001}(\varepsilon) = H(0) + \frac{dH}{dK_{001}}\left(\frac{1+\upsilon}{1-\upsilon}\right)(B + D\varepsilon)\varepsilon,$$

which can explain the experimental data for a suitable choice of magnetoelastic parameters.

*Acknowledgements*

We acknowledge fruitful discussions with Manh Duc Le, Jaehong Jong and Paul Graham. This work was supported by the French Research Agency (ANR) projects "Méloïc", "Nomilops", and "Multidolls," the European Research Council Advanced Grant "FEMMES" (contract n°267579) and the European Research Council Consolidator Grant "MINT" (contract n°615759).

Received: ((will be filled in by the editorial staff))
Revised: ((will be filled in by the editorial staff))
Published online: ((will be filled in by the editorial staff))

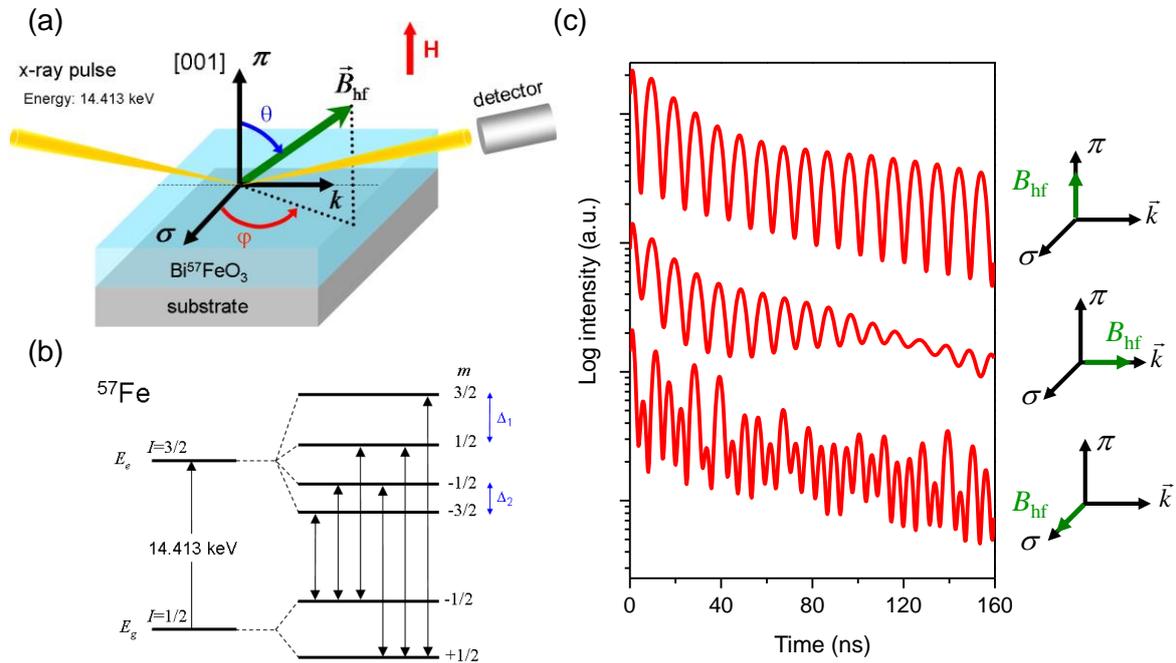

**Figure 1. Nuclear resonant scattering of synchrotron radiation from a Bi$^{57}$FeO$_3$ (BFO) thin film**. **(a)**, Geometry used in nuclear resonant scattering of synchrotron radiation from a thin film, with σ and π the linear polarization basis vectors and (θ,φ) the relative orientation of the magnetic hyperfine field $B_{hf}$ with respect to the incident wave vector *k*. **(b)**, Splitting of nuclear levels of $^{57}$Fe in the case of combined quadrupole electric and dipolar magnetic hyperfine interactions. **(c)**, Calculated nuclear time spectra for a BFO thin film with different orientations of $B_{hf}$.

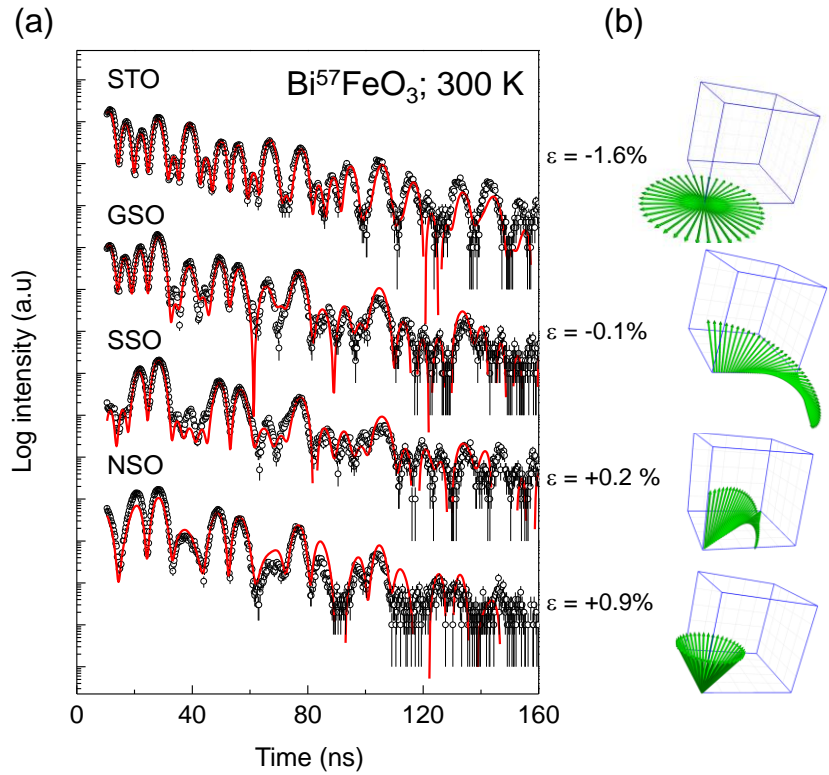

**Figure 2. Nuclear time spectra at room temperature of strained Bi$^{57}$FeO$_3$ thin films.** (a) The data show a systematic change of the beat pattern for BFO films epitaxially grown on different substrates (STO, GSO, SSO and NSO) spanning a strain range from compressive to tensile, indicating a strong dependence of the magnetic structure on strain. (b) Magnetic structure of the BFO films deduced from fits of the nuclear time spectra.

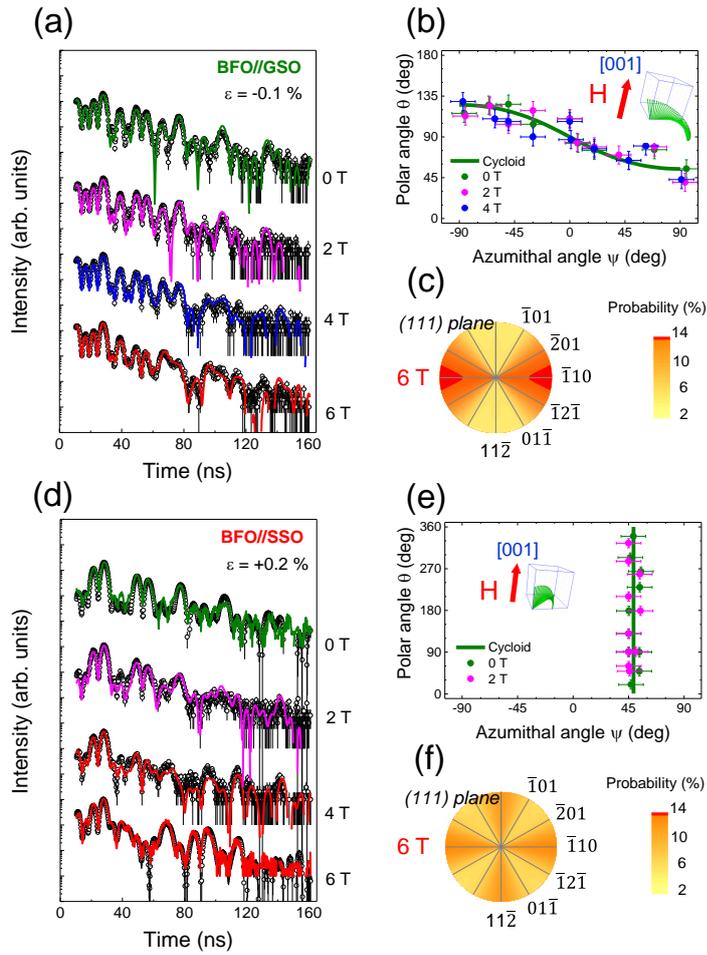

**Figure 3. Evolution of the spin cycloidal modulation under applied magnetic field.** (a),(d) Nuclear time spectra at 295 K of BFO//GSO and BFO//SSO thin films. (b),(e) Plots of the orientation of the hyperfine field (θ-φ) derived from the fit of the nuclear time spectra for various values of applied magnetic field. The insets show sketches of the magnetic structure for the two cycloidal arrangements below the critical field. (c),(f) Probability of hyperfine field direction in the (111)-type magnetic easy plane deduced from the fit of the NRS spectra at 6 T. The preferred orientations of the Fe spins are denoted by darker regions.

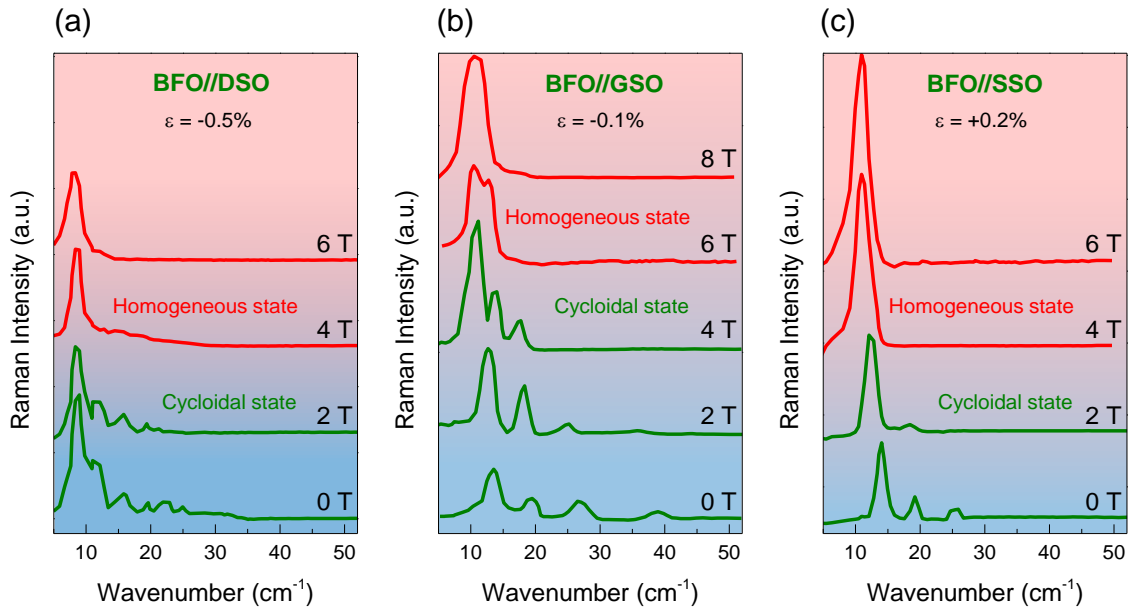

**Figure 4. Low-energy Raman fingerprint of the spin cycloid under magnetic field.**

Raman spectra at 295 K of (a) BFO//DSO; (b) BFO//GSO; and (c) BFO//SSO thin films upon application of a magnetic field normal to the film plane.

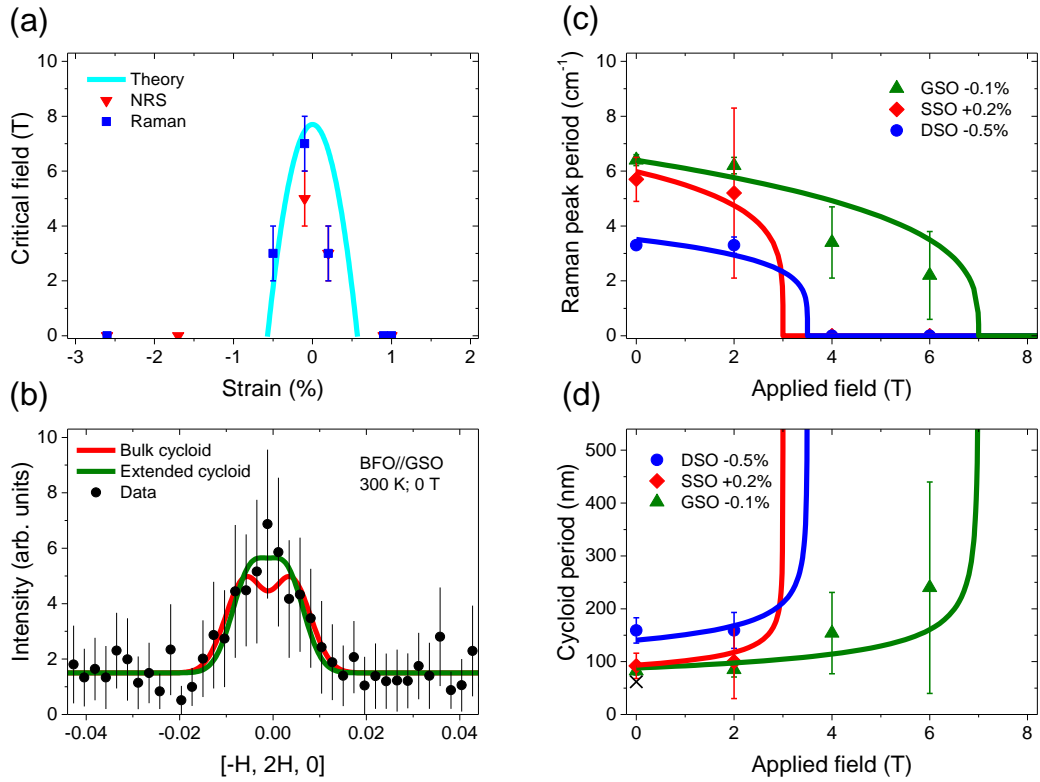

**Figure 5. Cycloid properties in BiFeO$_3$ thin films under magnetic field.** (a) Magnetic field – strain phase diagram of magnetic order in BiFeO$_3$: points represent experimental values, while the solid line demarcates the transition between cycloidal and homogeneous order for the type-1 (bulk-like) cycloid. (b) Diffracted neutron intensity integrated along the (-1,1,0) direction of the (½,½,½) magnetic reflection for BFO//GSO. The green line denotes a fit using a cycloid with period identical to the bulk (62 nm), while an extended period (82 nm) cycloid fit is shown with a red line. (c) Evolution with magnetic field of the peak period in the low energy Raman spectra from Fig. 4. The lines are guides to the eye. (d) Period of the cycloid (assuming a constant cyclon energy), derived from low-energy Raman spectra, showing that as the critical field is approached, the period is increased for all three strain levels. The lines are guides to the eye (see text).

**Supporting Information**

**Method for fitting nuclear resonant scattering (NRS) spectra**

NRS spectra corresponding to different magnetic structures in (001)-oriented BiFeO$_3$ (BFO) were calculated using the NRS package of the CONUSS program[41]. We considered various orientations of the magnetic hyperfine field, which are described by the polar coordinates $\theta$ (polar angle, "out-of-plane") and $\varphi$ (azimuthal angle, "in-plane") [Figure 1(a)]. To do this, we took into account different Fe configurations, corresponding to the same crystallographic site and with the same hyperfine parameters, but with different orientations of the hyperfine field, *i.e.* different values of $\theta$ and $\varphi$.

The case of uniaxial anisotropy corresponds to a single orientation of the hyperfine field $B_{hf}$, and therefore a single iron site. For a harmonic cycloidal modulation of the spins, we used a distribution of 10 orientations ($\theta$, $\varphi$) of $B_{hf}$, with the same probability (*i.e.* relative weight) for each iron site.

The direction of the electric field gradient (EFG) is described by the angles $\alpha$, $\beta$, and $\gamma$, which represent the angle subtended by the polarization vector and, respectively, the wave vector $\vec{k}$, the normal to the wave vector in the plane $\sigma$, and the normal to the film plane $\pi$. For BFO with polarization along [111], these angles are $\alpha = 35.26°$, $\beta = 90°$, and $\gamma = 54.74°$.

1. **Cycloidal spin distribution**

For a complex spin structure – such as a cycloidal modulation – the calculation of an NRS spectrum requires taking into account an anisotropy of the hyperfine interaction in the cycloidal plane, and then to calculate the values ($\theta$, $\varphi$) for each orientation of the hyperfine field $B_{hf}$.

The anisotropy of the hyperfine interaction was directly introduced in the program CONUSS by modulating the value of $B_{hf}$ as a function of its orientation to the principal axes of the EFG. **Figure S1** presents the variation in $B_{hf}$ used for our fits. We considered a variation of $B_{hf}$ over an interval [47.7 T, 49.8 T], comparable to the values used by Zalesskii et al.[44] to fit the asymmetry in the NMR spectra for $^{57}$Fe in bulk BFO.

The hyperfine field rotates inside the plane containing the main axis of the EFG [111] and the cycloid propagation direction (Figure S1). The resulting variation in the polar angle $\theta$ and azimuthal angle $\varphi$ can then be calculated using spherical trigonometry formulae. One can thus relate the angle $\theta$ to the angle $\theta_{cyc}$ using

$$\cos\theta = \cos\theta_{cyc} \times \cos(54.74°) + \sin\theta_{cyc} \times \sin(54.74°) \times \cos\Omega, \quad (4.2)$$

$$\cos\theta = \frac{1}{\sqrt{3}}\cos\theta_{cyc} + \sqrt{\frac{2}{3}}\sin\theta_{cyc}\cos\Omega, \quad (4.3)$$

where $\Omega$ denotes the angle between the cycloidal plane (in green in Figure S1) and the plane ($\pi$, [111]).

For type-1 and type-2 cycloids, $\Omega = 90°$ and $0°$, respectively, and equation (4.3) reduces as

$$\cos\theta = \frac{1}{\sqrt{3}}\cos\theta_{cyc} \quad \text{(type-1 cycloid)}, \quad (4.4a)$$

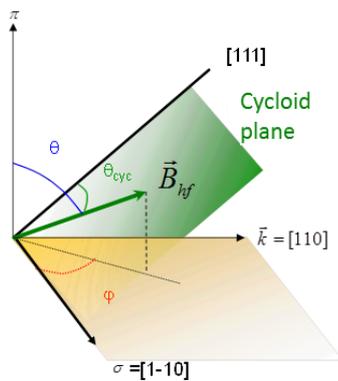
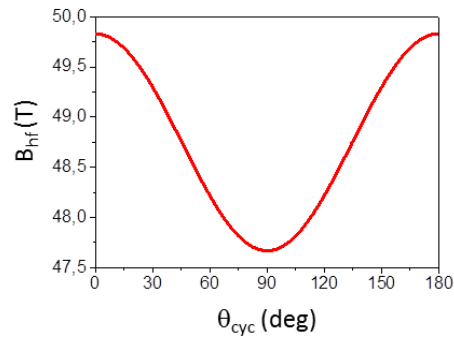

**Figure S1.** Angular dependence of $B_{hf}$ in the type-1 cycloidal plane defined by the angle $\theta_{cyc}$ relative to the direction [111]. The x-ray beam is assumed to be incident along the [110] direction.

$$\cos\theta = \cos(\theta_{cyc} - 54{,}74°) \qquad \text{(type-2 cycloid)}. \qquad (4.4b)$$

For the type-1 cycloid, we have

$$\sin\theta = \frac{\sqrt{\frac{2}{3}}\cos\theta_{cyc}}{\sin\varphi}, \qquad (4.5)$$

from which we obtain

$$\tan\theta = \frac{\sqrt{2}}{\sin\varphi} \qquad \text{(type-1 cycloid)}. \qquad (4.6)$$

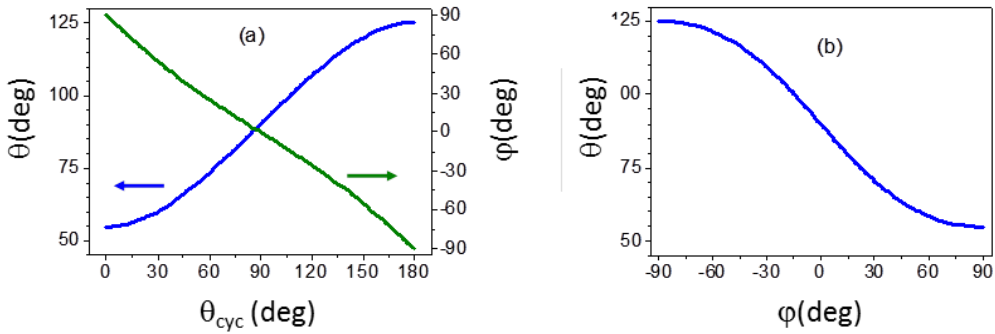

**Figure S2.** (a) Variation of the angles θ (blue curve) and φ (green curve) as a function of $\theta_{cyc}$, and (b) θ as a function of φ for the type-1 cycloid.

The dependence of the angles $\theta$ and $\varphi$ on the angle $\theta_{cyc}$, and that of $\theta$ as a function of $\varphi$, for the type-1 cycloid are presented in Figure S2. The calculated NRS spectrum using the angles obtained with Equations 4.5 and 4.6, and the values of the hyperfine field (shown in Figure S1), is presented in Figure S3(a). The spectrum thus obtained is the characteristic of the type-1 cycloid in BFO, that is, with a propagation vector along the direction < 1-10 >.

The type-2 cycloid is defined by a distribution of spins in a vertical plane, the azimuthal angle is therefore constant and equal to 45° if the incident beam is oriented along the [110] direction. The simulated NRS spectrum and the corresponding magnetic structure, which are characteristic of the type-2 cycloid, are presented in Figure S3(b). The spectrum has a form that is markedly different from that obtained using the type-1 cycloid spin distribution [Figure S3(a)], allowing us to distinguish the two spin structures without ambiguity.

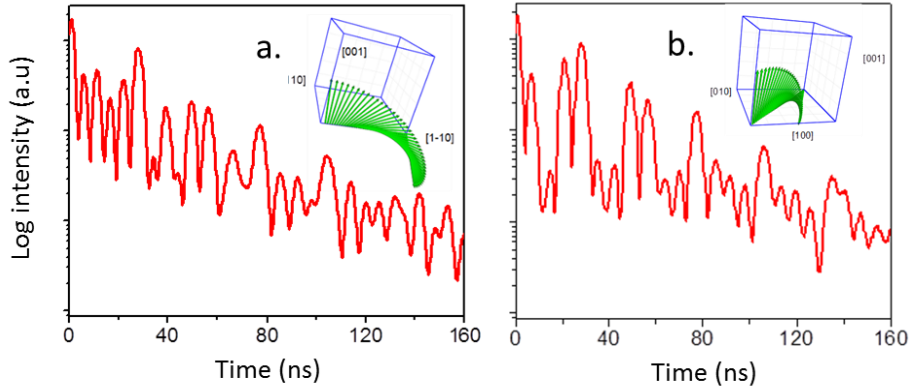

**Figure S3.** NRS spectra calculated for (a) type-1 and (b) type-2 cycloidal spin distributions in BFO.

2. **Distribution of spins in a (111) plane**

In the case of a spin distribution in a (111) easy-magnetic plane, we have

$$\cos\theta = \sqrt{\frac{2}{3}}\cos\theta_{cyc} \text{ and } \sin\theta = \frac{\sin\theta_{cyc}}{\cos(\varphi+45°)}, \qquad (4.12)$$

which yields

$$\cos^2\theta = \frac{\sin^2(\varphi+45°)}{\frac{3}{2}-\cos^2(\varphi+45°)}. \qquad (4.13)$$

The dependence of $\theta$ on $\varphi$ given by Equation (4.13), the resulting NRS spectrum, and the corresponding spin distribution, are plotted in Figure S4.

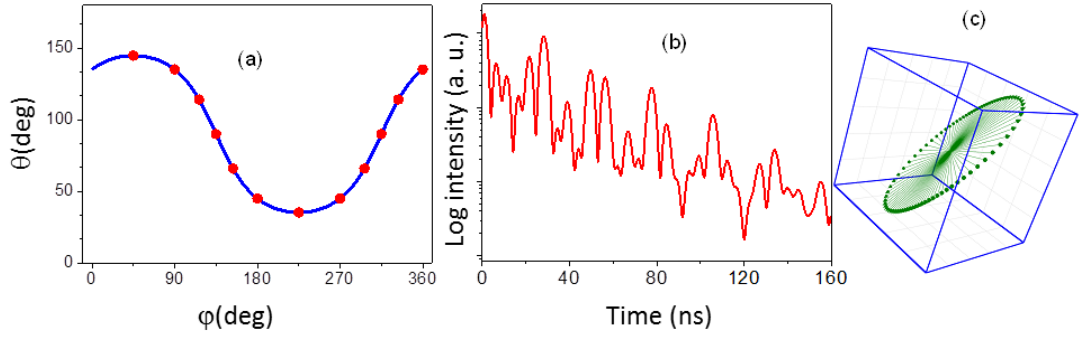

**Figure S4.** (a) Variation of the polar angle θ as a function of the azimuthal angle φ for a distribution of spins in a (111) easy plane. (b) Calculated NRS spectrum for spins in the (111) plane, and (c) the corresponding spin distribution.

**Calculation of magnetoelastic energy as a function of strain**

Using the Landau-Lifshitz formalism, as described in the experimental section, we can use the numerical dependence of the critical field [Figure 5(a)] obtained for the particular cycloid to obtain the dependence of the magnetoelastic energy on the elastic strain. Using the experimental data of the dependence of the critical field on strain for our BiFeO$_3$ films, the magnetoelastic energy for the type-1 cycloid is presented as a function of strain in Figure S5. Note that for films in which the cycloid is not present at 0 T (*i.e.* a homogeneous spin state) (strain magnitude |ε| > 0.75 %) we estimate a *lower limit* for the magnetoelastic energy.

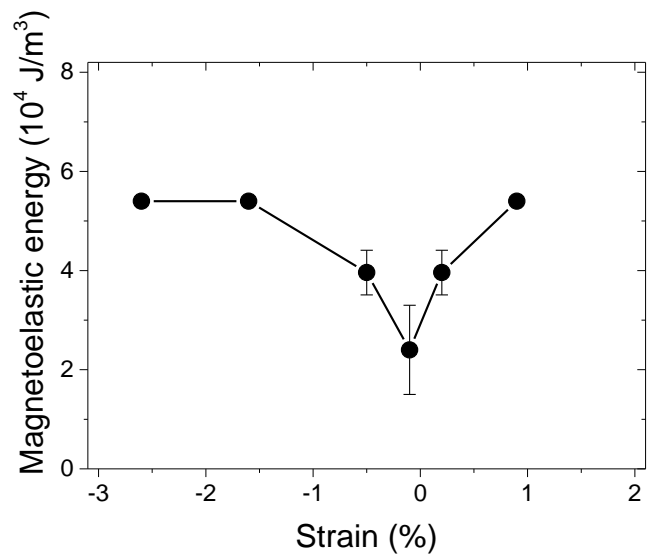

**Figure S5.** Calculated magnetoelastic energy for BiFeO$_3$ films as a function of epitaxial strain, based on the critical magnetic field for the cycloid → homogeneous order transition.